# Solitons and Instantons in the inverse $\varphi^4$ theory of critical phenomena. A proposal for the violation of causality of tachyonic field.

Y. Contoyiannis[1], P. Papadopoulos[1], M. Kampitakis[2], M.P. Hanias[3], S.G. Stavrinides[4]

[1]Department of Electric-Electronics Engineering, University of West Attica, Athens, Greece. (yiaconto@uniwa.gr, ppapadopoulos@uniwa.gr,
[2]Major Network Installations Dept, Hellenic Electricity Distribution Network Operator SA, Athens, Greece, (m.kampitakis@deddie.gr)
[3]Physics Department, International Hellenic University, Kavala, Greece. (mhanias@physics.ihu.gr)
[4]School of Science and Technology, International Hellenic University, Thessaloniki, Greece. (s.stavrinides@ihu.edu.gr)

**Abstract:** We investigate about the existence of static solitons solutions in the inverse G-L free energy in phase transitions of $\varphi^4$ theory. We calculate all the characteristics of these solitons, like localized structure, finite energy and mass. We show that solitons appears in spontaneous symmetry breaking (SSB) phenomenon only if the critical point is in excited state. When the time is introduced we shown that the static solitons of SSB remain as solitons in Minkowski space-time while the static solitons of inverse G-L free energy converted to instantons in Euclidean space. The violation of causality which appears in the tachyonic field could be faced passing from Minkowski solitons to Euclidean instantons.
.

## 1. Introduction

The main important issue in critical phenomena and phase transition is the critical point. In the symmetric phase the critical point has a role of saddle point in a potential function which according to Landau theory [1] has polynomial form in order parameter $\varphi$ which is invariant under $\varphi \to -\varphi$ :

$$U(\varphi) = \tfrac{1}{2}r_o\varphi^2 + \tfrac{1}{4}u_o\varphi^4 + \tfrac{1}{6}c_o\varphi^6 + \tfrac{1}{8}s_o\varphi^8 + .. (1)$$

where the coefficients in eq.1 are all positive.

In above free energy the terms up to $\varphi^6$ are enough to describes the second and the first order phase transitions. So we have :

- The free energy $U(\varphi) = \tfrac{1}{2}r_o\varphi^2 + \tfrac{1}{4}u_o\varphi^4$ describes the symmetric phase of

- a second order phase transition
- The free energy $U(\varphi) = \frac{1}{2}r_o\varphi^2 + \frac{1}{4}u_o\varphi^4 + \frac{1}{6}c_o\varphi^6$ describes the symmetric phase of a first order phase transition

  The symmetric breaking in $\varphi^4$ theory is accomplished when $r_o$ changes sign and became negative while $u_o > 0$. Then the critical point changes from stable to unstable and two new stable state are created.

  The symmetric breaking in $\varphi^6$ theory is accomplished when $u_o$ changes sign and became negative while $r_o > 0, c_o > 0$. Then the critical point remain as ground state while simultaneously two new ground state are created.

Solitons are finite-energy, localized solutions to the equations of motion that, after collisions, retain their shape. Solitons solutions has been found in the past for the above theories in the phase of broken symmetry. So, in effective form of $\varphi^4$ theory have been found [2] solitons:

$$U_{\varphi^4}(x) = -\frac{1}{2}|r_o|\varphi^2 + \frac{1}{4}u_o\varphi^4 + \frac{r_o^2}{4u_o} \quad (2)$$

In eq. (2) has been added a constant term. This mean that the critical point does not is in ground state but in a excited state with potential energy $\frac{r_o^2}{4u_o}$. These solitons are kink type.

For the $\varphi^6$ theory has been found [3] solitons in its effective form:

$$U_{\varphi^6} = \frac{1}{2}r_o\varphi^2 - \frac{1}{4}|u_0|\varphi^4 + \frac{1}{6}c_o\varphi^6 - f_o \quad (3)$$

In eq. (3) has been added a constant term $f_o$ determined boundary conditions imposed on system. In the regime of values where the degeneration of the order parameter appears ( first order phase transition) the solitons are half-kink type.

The scope of our work is to investigate soliton solution in an "intermediate" theory which is the inverse G-L free energy without addition of constant term. This research will reveals the existence of solitons type pulse . In the domain of time these solutions are instantons .

## 2. The inverse G-L free energy

The inverse G-L free energy is given as :

$$U_{in}(\varphi) = \frac{1}{2}r_o\varphi^2 - \frac{1}{4}|u_0|\varphi^4 \quad (4)$$

This theory which expressed through eq. (4) is an "intermediate" theory because consist from self-interactions up to $\varphi^4$, but the change of sign has been done in term of $\varphi^4$ like the $\varphi^6$ theory. The $U_{in}(\varphi)$ is the inversion of $U_{\varphi^4}(x)$ ( without the constant term) , namely , $U_{in}(\varphi) = -U_{\varphi^4}(x)$. In the other hand the potential (4) could be consider as a weak $\varphi^6$ theory where in eq.3 $c \to 0$. In fig 1 this inversion is shown. We see, with green line, the symmetric phase $U(\varphi) = \frac{1}{2}r_o\varphi^2 + \frac{1}{4}u_o\varphi^4$ ($r_o > 0, u_o > 0$) . The SSB phenomenon ($r_o < 0, u_o > 0$) presented with blue line where the two stable point are demonstrated , and the inverse phenomenon (($r_o > 0, u_o < 0$) presented with red line where in the position of stable points appears two unstable points.

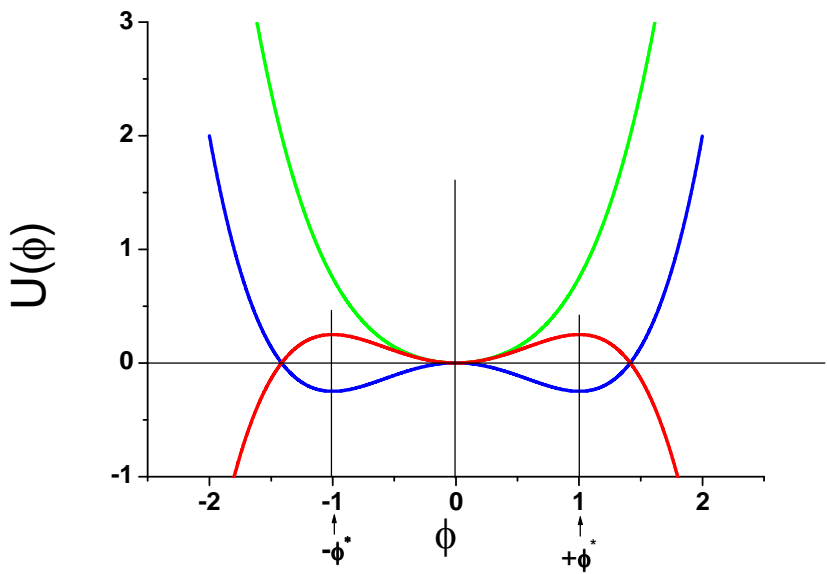

*Fig. 1 With green the symmetric phase is shown. With blue and red the SSB and the inverse G-L free energy respectively are presented , as describes in text. The maxima and minima in red, blue line respectively are marked at* $\varphi^* = \pm\sqrt{\frac{r_o}{u_o}}$, *with* $r_o = u_o = 1$.

The soliton solution in SSB (blue line) express the passage between the two local energy minima. This soliton $\varphi(x)$ is a kink soliton which describes this passage. The soliton in inverse G-L free energy ( if this exist) express the passage between the two local energy maxima (red line). The procedure which we follow to investigate about solitons is to face the problem for static solutions and after this we introduce the time . In this way we take the soliton solution $\varphi(x, t) = f(x - vt)$.

## 3. Static Solitons (solitons in space) of inverse G-L free energy.

We follow the standard procedure for the investigation about the solitons solutions [2].

Starting from Euler-Lagrange equation for classical field $\varphi(x,t)$ we have:

$$\frac{\partial^2}{\partial^2 t}\varphi(x,t) - \frac{\partial^2}{\partial^2 x}\varphi(x,t) = -\frac{\partial}{\partial \varphi}U(\varphi) \quad (5)$$

For static solutions we have from eq.(5):

$$\frac{\partial^2}{\partial^2 x}\varphi(x) = \frac{\partial}{\partial \varphi}U(\varphi) \quad (6)$$

If we multiply the eq.(6) with $\frac{\partial \varphi}{\partial x}$, we have:

$$\frac{\partial^2 \varphi}{\partial^2 x} \cdot \frac{\partial \varphi}{\partial x} = \frac{\partial U(\varphi)}{\partial \varphi} \cdot \frac{\partial \varphi}{\partial x} \quad (7)$$

The eq.(7) can be integrated over x, yielding:

$$\int \frac{\partial}{\partial x}\left(\frac{\partial \varphi}{\partial x}\right) \cdot \frac{\partial \varphi}{\partial x} dx = \int \frac{\partial U(\varphi)}{\partial \varphi} \cdot \frac{d\varphi}{dx} dx = U(\varphi) \quad (8)$$

The quantity $\int \frac{\partial}{\partial x}\left(\frac{\partial \varphi}{\partial x}\right) \cdot \frac{\partial \varphi}{\partial x} dx = \left(\frac{\partial \varphi}{\partial x}\right)^2 - \int \frac{\partial}{\partial x}\left(\frac{\partial \varphi}{\partial x}\right) \cdot \frac{\partial \varphi}{\partial x} dx \Rightarrow 2\int \frac{\partial}{\partial x}\left(\frac{\partial \varphi}{\partial x}\right) \cdot \frac{\partial \varphi}{\partial x} dx = \left(\frac{\partial \varphi}{\partial x}\right)^2 \Rightarrow \int \frac{\partial}{\partial x}\left(\frac{\partial \varphi}{\partial x}\right) \cdot \frac{\partial \varphi}{\partial x} dx = \frac{1}{2}\left(\frac{\partial \varphi}{\partial x}\right)^2 \quad (9)$

From eqs. (8), (9) we take:

$$U(\varphi) = \frac{1}{2}\left(\frac{\partial \varphi}{\partial x}\right)^2 \quad (10)$$

This equation can be integrated once again to yield:

$$x - x_0 = \pm \int_{\varphi(x_0)}^{\varphi(x)} \frac{d\bar{\varphi}}{\sqrt{2U(\bar{\varphi})}} \quad (11)$$

Putting in eq.(11) the $U(\varphi) = U_{in}(\varphi)$ from eq.(4) we take:

$$x - x_0 = \pm \int_{\varphi(x_0)}^{\varphi(x)} \frac{d\bar{\varphi}}{\sqrt{r_o \bar{\varphi}^2 - 0.5 u_o \bar{\varphi}^4}} \quad (12)$$

$$x - x_0 = \pm I$$

Where $\quad I = -\int_{\varphi(x)}^{\varphi(x_0)} \frac{d\bar{\varphi}}{\sqrt{r_o \bar{\varphi}^2 - 0.5 u_o \bar{\varphi}^4}} = -\frac{1}{\sqrt{0.5 u_o}} \int_{\varphi(x)}^{\varphi(x_0)} \frac{d\bar{\varphi}}{\bar{\varphi}\sqrt{\frac{r_o}{0.5 u_o} - \bar{\varphi}^4}}$

Using the integral [4] $\int \frac{dz}{z\sqrt{a^2 - z^2}} = -\frac{1}{a} arcsech\left(\frac{z}{a}\right)$ we take that:

$$x - x_0 = \pm(\frac{1}{\sqrt{0.5u_o}} \frac{1}{\sqrt{\frac{r_0}{0.5u_o}}} (\operatorname{arcsech}\left(\frac{\varphi(x_0)}{\sqrt{\frac{r_0}{0.5u_o}}}\right) - \operatorname{arcsech}\left(\frac{\varphi(x)}{\sqrt{\frac{r_0}{0.5u_o}}}\right))$$

Due to $\varphi(x_0)$ is the upper integration limit we have

$$\operatorname{arcsech}\left(\frac{\varphi(x_0)}{\sqrt{\frac{r_0}{0.5u_o}}}\right) = \operatorname{arcsech}(1) = 0$$

So, $x - x_0 = \mp \frac{1}{\sqrt{r_0}} \operatorname{arcsech}\left(\frac{\varphi(x)}{\sqrt{\frac{r_0}{0.5u_o}}}\right)$. Inverting, we then find:

$$\varphi(x) = \pm \sqrt{\frac{r_o}{u_o}} \frac{\sqrt{2}}{\cosh((x-x_o)\sqrt{r_o})} \quad (13)$$

In fig.2a the soliton $\varphi(x)$ is shown. Similar solitons described by eq. (13) appears in QCD phase transitions [5].

The energy density is :

$$\mathcal{E} = T + U = \frac{1}{2}(\frac{\partial \varphi}{\partial x})^2 + U \ (14)$$

Where T the kinetic density term. From eqs (10,14) we take that :

$$\mathcal{E} = (\frac{\partial \varphi}{\partial x})^2 \ (15)$$

From eq.(13) we estimate the $\frac{\partial \varphi}{\partial x}$ :

$$\frac{\partial \varphi}{\partial x} = -\frac{r_o \sqrt{2}}{\sqrt{u_o}} \operatorname{sech}((x - x_o)\sqrt{r_o}) \tanh((x - x_o)\sqrt{r_o}) \ (16)$$

We substitute the eq.(16) in eq. (15) and we take :

$$\mathcal{E}(x) = \frac{2r_o^2}{u_o} \operatorname{sech}^2(x - x_o)\sqrt{r_o}) \tanh^2((x - x_o)\sqrt{r_o}) \ (17)$$

In fig 2b. the energy density is shown.

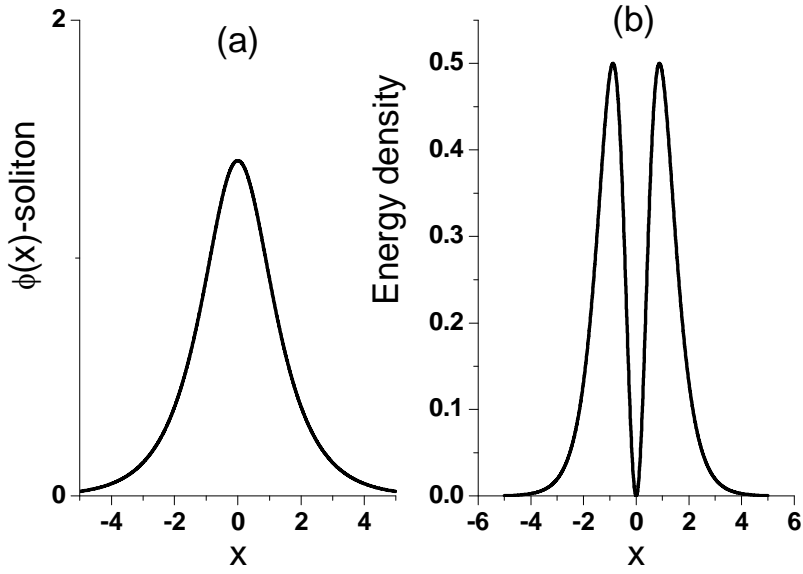

*Fig.2 (a) The field of soliton for the inverse G-L free energy. This soliton is type pulse. (b) The energy density of the of soliton for the inverse G-L free energy. The parameters have the values $r_o = u_o = 1$, $x_o = 0$.*

We see in Fig. 2 the existence of a pulse soliton which is characterized by a localized structure in space and finite energy.

The mass of soliton is given by the integral over the energy density :

$$M = \int_{-\infty}^{\infty} \mathcal{E}(x)dx = \frac{2r_o^2}{u_o}\int_{-\infty}^{\infty} \text{sech}^2(x-x_o)\sqrt{r_o})\tanh^2((x-x_o)\sqrt{r_o})dx = \frac{4r_o^2}{u_o}\int_0^{\infty}\frac{\sinh^2((x-x_o)\sqrt{r_o})}{\cosh^4((x-x_o)\sqrt{r_o})}dx$$

Using the integral $\int_0^{\infty}\frac{\sinh^{\mu}z}{\cosh^{\nu}z}dz = \frac{1}{2}B(\frac{\mu+1}{2},\frac{\nu-\mu}{2})$ [6], we find:

$$M = \frac{2r_o^{3/2}}{u_o}B\left(\frac{3}{2},1\right) = \frac{2r_o^{3/2}}{u_o}\frac{\Gamma(\frac{3}{2})\Gamma(1)}{\Gamma(\frac{5}{2})} = \frac{2r_o^{3/2}}{u_o}\frac{0.5\sqrt{\pi}}{\frac{3}{4}\sqrt{\pi}} = \frac{4r_o^{3/2}}{3u_o} \quad (18)$$

## 4. Static Solitons (solitons in space) of SSB.

### 3a. Critical point in ground state ($\varphi = 0, U = 0$)

This is the SSB case when the critical point is the state ($\varphi = 0, U = 0$), namely the blue line. Following the steps of section 3 we take that for field :

$$\varphi(x) = \pm\sqrt{\frac{r_o}{u_o}}\frac{\sqrt{2}}{\cos((x-x_o)\sqrt{r_o})} \quad (19)$$

And for energy density :

$$\mathcal{E}(x) = \frac{2r_o^2}{u_o}\sec^2((x-x_o)\sqrt{r_o})\tan^2((x-x_o)\sqrt{r_o})\quad (20)$$

In fig. 3 the field $\varphi(x)$, $\mathcal{E}(x)$ are shown.

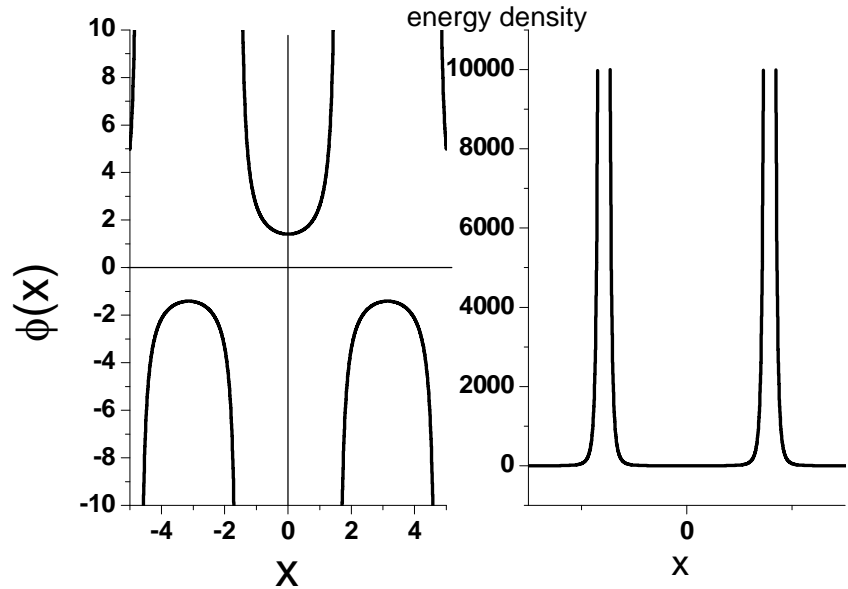

**Fig.3 (a) The field $\varphi(x)$ is not localized in space (b) The energy density is not finite**

From fig.3 results that soliton solution for this case no exist because there are not localized structures in space nor finite energy. We examine if the reason for something like this is that the minima of blue line has negative value for the potential $U = -\frac{r_o^2}{4u_o}$.

3b. **Critical point in excited state ($\varphi = 0, U = \frac{r_o^2}{4u_o}$)**

In this case the SSB potential is given by eq.2
This case has faced [2] and the results are:

$$\varphi(x) = \pm\sqrt{\frac{r_o}{u_o}}\tanh\left[\left(\sqrt{\frac{r_o}{2}}\right)(x-x_o)\right]\quad (21)$$

And

$$\mathcal{E}(x) = \frac{r_o}{2u_0} \operatorname{sech}^4[\left(\sqrt{\frac{r_o}{2}}\right)(x - x_o)] \quad (22)$$

In fig. 4 the field $\varphi(x), \mathcal{E}(x)$ are shown.

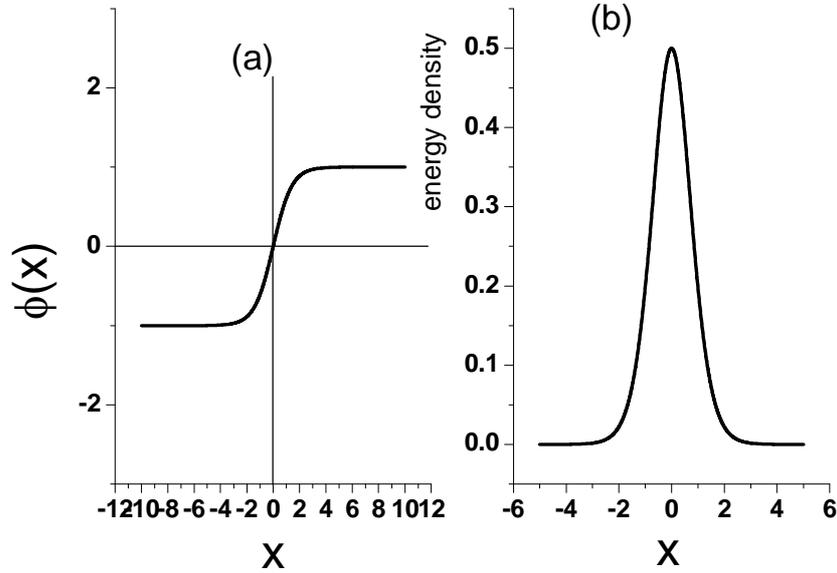

*Fig.4 (a) soliton in SSB is localized in space (b) Energy density in SSB is finite.*

It is obvious the appearance of a kink soliton now. We see a localized structure between the asymptotic field values $\varphi(|x| = \infty) = \pm\sqrt{\frac{r_o}{u_o}}$ and we found finite energy.

### 5. Solitons in space-time

Before we proceeds to introduction of time in soliton solutions we investigate the origin of the passage between vacua. More generally the passage from a vacuum to an other vacuum is a tunneling effect. The transmission amplitude is given from semi-classical approximation WKB where in the time is written as :

$$\Gamma(t) = \exp\left[(-\frac{1}{\hbar}\int dt[2\sqrt{V - E})\right] \quad (23)$$

In the case of SSB ( blue line) the barrier $V$ is greater than the E of soliton –particle and so the quantity $V - E > 0$. The problem of the passage is solved in the in the real time $t$ of Minkowski space. This space is appropriate because is compatible to special relativistic theory in order to includes relativistics theories.

The case of inverse G-L free energy ( red line) belongs to the category where does not exist barrier ($V = 0$) or the barrier is smaller than the energy ($V - E < 0$). These are the cases of the classical motion , for example in the case of red line a ball rolling down one hill and arriving at the other hill. Then the eq. 23 is written:

$$\Gamma(t) = \exp\left[\left(-\frac{1}{\hbar}\int d(it)[2\sqrt{E-V}\right.\right. = \exp\left[\left(-\frac{1}{\hbar}\int d\tau[2\sqrt{E-V}\right]\right. \quad (24)$$

Where $\tau = it$ is the Euclidean time.

We conclude that the SSB solitons lives in Minkowski space- time while the inverse G-L free energy solitons lives in Euclidean space-time. The solitons in Euclidean space-time are called instantons.

So the SSB soliton moving at velocity $v$ taking into consideration the relativity is written as :

$$\varphi(x,t) = \pm\sqrt{\frac{r_o}{u_o}} \tanh\left[\left(\sqrt{\frac{r_o}{2}}\right)\frac{(x-x_o)-vt}{\sqrt{1-(\frac{v}{c})^2}}\right] \quad (24)$$

While for instanton moving at velocity $v$ is written as :

$$\varphi(x,\tau) = \pm\sqrt{\frac{r_o}{u_o}} \frac{\sqrt{2}}{\cosh\left((x-x_o) - v\tau\right)\sqrt{r_o})} \quad (25)$$

The boosting transformations for the Euclidean space is a controversial issue and so we don't write a equation like (24) in the case of instantons.

### 6. What is the instantons ?

Instantons connect the two unstable "vacua" of the inverse G-L free energy of red line in fig. 2. Solitons-instantons describes the solution of field equations as $\vec{x}$ ($x_1, x_2, x_3$) evolves from $-\infty$ to $+\infty$. Moreover the instantons describes the solution of field equations as the $x_4 \equiv \tau$ , in Euclidean space, evolves from $-\infty$ to $+\infty$. These unstable vacua are different between them. This is imposed from their Euclidean origin because according to semiclassical approach the instantons must represent different paths in the path integral formalism around a classical solution. As it is proved [2] each vacuum is characterized by a winding number $n \in Z$. So, the instanton impose a mapping between the different vacua as $\| n \rangle \to \| n+1 \rangle$. This issue has topological character as a mapping $S_3 \to S_3$ , where $S_3$ is the boundary of $E_4$ euclidean space at $\tau \to \pm\infty$ [7]. This topological character does not exist to solitons where the vacua in SSB (green line) are equivalents. The mapping $\| n \rangle \to \| n+1 \rangle$ between the different vacua impose a "time" array in imaginary time $\tau$ as the time array in the real world.

Field Theories where topological solutions instanton exist are the pure gauge theories in Quantum Field Theory.

## 7. The tachyonic field

It is known that in field theories utilizing the mass as a control parameter, like in the case of free energy of the Higgs mechanism, the $\varphi^2$ coefficient is the mass $m^2$ [2]. Then during the spontaneous symmetry breaking, the ground state moves from the critical point to new states. This happens through a change in the sign of $m^2$ which becomes negative; thus the mass becomes an imaginary magnitude. This field is a tachyon [2], since according to the relativistic theory it moves faster than the light. The above is a characteristic case where violation of causality takes place. In general, the phenomenon of spontaneous symmetry breaking, which is closely related to tachyon condensation, plays an important role in many aspects of theoretical physics, including the Ginzburg–Landau [8] and BCS theories of superconductivity. The soliton solution (24) of $\varphi^4$ SSB could describe the tachyonic field if the relativistic problem could be faced. According to metioned above there is a way to avoid this problem through the instanton solution (25) which refer to inverse $\varphi^4$ SSB. Indeed, the soliton (24) lives to the real world of Minkowski space where violation of causality takes place when a particle exceeds the light speed while the instanton (25) live in the Euclidean space where such a violation of causality does not exist. But, as we saw previously, the instantons is rather a topological object than a particle. A space where such topological objects expressed through mapping $\parallel n \rangle \to \parallel n+1 \rangle$ is the classical phase space. Therefore, we could study the instantons field (25) instead tachyonic field, in the phase space which is produced from Poincare maps generated by inverse G-L potential. This is the object of a further work.

## *8. Conclusions*

In this work we have shown that the inverse G-L free energy in $\varphi^4$ theory has instanton solutions in the Euclidean space. In static version solitons appears with all properties of solitons like the constant structure of a particle with energy and mass. We also showed that in SSB the solitons appears only if the critical point is in excited state. If we added the time dimension in the space then the instanton, in contrary to solitons in Minkowski space, become a topological object expressed through a mapping $S_3 \to S_3$, where $S_3$ is the boundary of $E_4$ euclidean space at $\tau \to \pm\infty$ [7]. An ideal space for mapping is the classical phase space. In this framework we could study the instanton field given from (25) instead the tachyonic-soliton field given from (24) in order to we avoid the serious problems which emerges from the violation of causality.